\documentstyle[12pt,psfig]{article}
\def\refe#1{(\ref{#1})}

\def\ie{{\it i.e.}}

\def\LEP2{{LEPII}}
\def\mg{$m_{3/2}$}
\def\th{$\theta$}
\def\ms{$M_{\scriptscriptstyle String}$}
\def\mp{M_{\scriptscriptstyle Planck}}
\def\rad{rad.}

\def\npb#1#2#3{    {\it Nucl. Phys. }{\bf B #1} (19#2) #3}

\def\plb#1#2#3{    {\it Phys. Lett. }{\bf B #1} (19#2) #3}

\def\prep#1#2#3{   {\it Phys. Rep. }{\bf #1} (19#2) #3}

\def\rnc#1#2#3{    {\it Riv. Nuovo Cim. }{\bf #1} (19#2) #3}

\def\ibid#1#2#3{   {\it ibid. }{\bf #1} (19#2) #3}

\begin{document}
\begin{titlepage}
\vspace*{-1.5cm}
\begin{center}

\hfill IC/95/373
\\[1ex]  \hfill November, 1995

\vspace{1ex}
{\Large \bf \ Low Energy Implications 
\\[1ex]of Minimal Superstring Unification}

\vspace{2ex}
{\bf Shaaban Khalil}$^{a,b}$, {\bf Antonio Masiero}$^{c,d}$ {\bf and  
Francesco Vissani}$^{a,e}$\\
{\it
\vspace{1ex} a) International Centre For Theoretical Physics, ICTP,
via Costiera 11, I-34100,
Trieste, Italy.}\\
{\it
\vspace{1ex} b) Ain Shams University, 
Faculty of Science, 
Department of 
Mathematics, 
Cairo, Egypt.}\\
{\it
\vspace{1ex} c) Dipartimento di Fisica, Universit\`{a} di Perugia.}\\ 
{\it
\vspace{1ex} d) INFN, Sezione di Perugia, 
Via Pascoli, 
I-06100 Perugia, Italy.}\\
{\it
\vspace{1ex} e) INFN, Sezione di Trieste, c/o 
SISSA, Trieste, Italy.}\\

\vspace{3ex}
{ABSTRACT}
\end{center}

\begin{quotation}
  We study the phenomenological implications of effective supergravities
based on string vacua with spontaneously broken N=1 supersymmetry by
dilaton and moduli $F$-terms. 
We further require Minimal String Unification, namely that 
large string threshold corrections 
ensure the correct unification of the gauge couplings 
at the grand unification scale. The whole supersymmetric
mass spectrum turns out to be determined in terms of only two 
independent parameters, the dilaton-moduli mixing angle and the 
gravitino mass. In particular we discuss the region of the parameter
space where at least one superpartner is ``visible" at \LEP2. We find 
that the most likely candidates are the scalar partner of the right-handed 
electron and the lightest chargino, with interesting correlations between
their masses and with the mass of the lightest higgs. 
We show how discovering SUSY
particles at \LEP2\ might rather sharply discriminate between scenarios 
with pure dilaton SUSY breaking and mixed dilaton-moduli breaking.
\end{quotation}
\end{titlepage}
\vfill\eject

The stunning experimental confirmations of the Standard Model (SM) that 
have kept accumulating along these last years make it mandatory for any 
new physics to exactly reproduce the SM at the Fermi scale.\par

Given that the two major open questions of SM concern the
incorporation of gravity 
among elementary interactions and the origin or naturalness of the 
electroweak scale $(M_W \ll \mp)$, it is likely that this new physics 
might be based on a locally supersymmetric quantum field theory that 
contains the SM, gravity and, may be, some additional interactions in 
which known particles do not take part. The hierarchy of mass scales may 
then naturally result from the spontaneous breaking of local 
supersymmetry above $M_W$ by some non-perturbative mechanism. The 
decoupling of this mechanism at low energy would imply that the new 
physics at the Fermi scale should be describable by an effective 
lagrangian with an N=1 global supersymmetry (SUSY) explicitly broken by a 
set of soft terms~\cite{Nilles}. Clearly, these terms as seen from an 
$\mp$ point of  view, are entirely calculable in terms of the 
supergravity couplings.

On the other hand, the non-renormalizability of supergravity strongly 
favours the view that supergravity itself has ultimately to be 
considered as an effective theory valid only at $E \leq \mp$. The best 
candidate we have for a description of physics at or above $\mp$ is a 
heterotic superstring  theory. At the perturbative level, it possesses a 
large class of vacua that lead to effective N=1 supergravities below 
$\mp$, however the  implementation of the above mentioned programme to 
finally derive an  effective theory that at $M_W$ reproduces the SM is 
still far from being realized. The major obstacle is the still large 
ignorance of how to  handle the crucial non-perturbative properties of 
string theory. Most  believe that the non-perturbative breakdown of SUSY 
or the selection of the true string vacuum or the determination of the 
gauge couplings indeed result   from ``stringy" 
mechanism whose non-perturbative nature prevents us from a deeper 
comprehension.

This major difficulty has prompted several authors to parametrize the 
effects of this unknown non-perturbative physics into a set of arbitrary 
parameters of the low energy effective theory. Such a set comprises 
couplings which are calculable in string perturbation theory and 
couplings which genuinely depend on the non-perturbative aspects. 
Remarkably enough, even with such a general parameterization, several 
features which are common to the whole class of low energy effective 
supergravities emerge. In the work of Kaplunovski and 
Louis~\cite{Kaplunovsky} 
along these lines, the properties of non-perturbative couplings were 
constrained making some rather general assumption on the non-perturbative 
dynamics\footnote{The analysis of ref.\ \cite{Kaplunovsky} finds
its ground in previous extensive work on gaugino condensation
and duality-invariant effective lagrangians
\cite{Dine}.}.
In particular, it was assumed that the flatness of moduli and 
dilaton directions of the effective potential was lifted by such 
non-perturbative dynamics and that SUSY breaking arises from the 
non-vanishing vacuum expectation values (VEV) 
of the $F$-terms of the moduli $T_i$, and/or dilaton 
$S$ supermultiplets. We follow here the approach of Brignole, 
Iba\~{n}ez and 
Mu\~{n}oz where local SUSY breaking with vanishing 
cosmological constant is assumed to be saturated by the dilaton and 
moduli auxiliary fields. 
Within a specified compactification scheme the soft terms become function 
of the gravitino mass and of the so-called goldstino angle, \ie\ the 
angle which accounts for 
the relative magnitude of the $T_i$ and $S$ $F$-terms VEV's in the SUSY 
breaking~\cite{Ibanez}.

In this letter we make use of the above general frame to study the 
implications of effective supergravities which emerge in the low-energy 
limit of superstring theory for \LEP2\ physics. In particular, we will 
discuss the three distinct situations which can be encountered with 
dilaton $ \langle F^S \rangle$ dominance $( \langle F^S \rangle \gg 
\langle F^T \rangle )$, moduli  $\langle F^T \rangle$ 
dominance $( \langle F^T \rangle \gg \langle F^S \rangle )$ or comparable 
role of dilaton and moduli 
$( \langle F^S \rangle \simeq \langle F^T \rangle )$ in SUSY breaking. 
The phenomenological implications of SUSY breaking solely due to the 
dilaton $F$-term were discussed in~\cite{Barbieri}. A 
further specification that we adopt for the 
class of superstring theories under analysis is related to the well-known 
problem of gauge couplings unification. We assume the so-called Minimal 
Superstring Unification~\cite{Ross}, \ie\ that the only light particles 
with SM gauge 
couplings are just those of the minimal SUSY SM (MSSM) and no partial
(field theoretical) unification occurs below the string scale. Then one has 
to rely on the string threshold contribution to cover the gap between the 
unification scale and the string scale. In orbifold 
compactification 
this was shown to be possible under rather constrained circumstances, \ie\ 
with a particular choice of the modular weights of the matter fields. 
This point will be further discussed below.

Assuming the presence of one dominant modulus $T$, the orbifold 
compactification and target space modular invariance, and the minimal 
matter and Higgs  content 
($Q,$ $U^c,$ $D^c,$ $L,$ $E^c$ and $H_1$ and $H_2$ superfields), 
the soft breaking terms can be expressed in terms of the modular weights 
and only two other
parameters at the compactification scale: the gravitino mass \mg\ and 
the goldstino angle \th, defined by
$\tan\theta={\langle F^T \rangle  / \langle F^S \rangle}$ 
($\theta=\pi/2$ corresponds to pure 
dilaton scenarios).

The scalar masses squared $m_i^2$ read \cite{Ibanez}: 
\begin{equation}
m_i^2= m^2_{3/2}\ (1+n_i \cos^2 \theta),
\label{scalar}
\end{equation}
where $n_i$ are integer numbers, known as modular weights.
A possible way to constrain the modular weights is provided by the demand 
to have minimal string unification. As discussed above this constraint 
entails a severe limitation on the available values of the modular 
weights. Some time ago Iba\~{n}ez, 
Lust and Ross~\cite{Ross} showed that assuming 
generation independence for the modular weights as well as $-3\leq n_i 
\leq 1$, minimal string unification could be achieved for 
\begin{equation}
\begin{array}{lll}
n_{L}=-3, & n_{E^c}=-3,&  \\ \nonumber
n_{Q}=-1, & n_{D^c}=-1, & n_{U^c}=-2, \\ \nonumber
n_{H_1}=-2, & n_{H_2}=-3,& 
\label{modular}
\end{array}
\end{equation}
or also the same as above with the replacement 
$n_{H_1} \leftrightarrow n_{H_2}.$

Obviously, the choice of the values of the modular weights has a major 
impact on the phenomenological implications that we wish to study here. 
It might turn out that the $M_X$---\ms\ discrepancy will be finally 
overcome in schemes (intermediate GUT, extra light states between  $M_X$ 
and  \ms, {\em etc.}) other than in the minimal string unification approach 
that we follow here. In any case we find it interesting to adopt this 
promising solution taking it seriously and trying to fully explore its 
impact on the coming \LEP2\ physics.

In the soft sector of the trilinear scalar couplings we focus only on the 
$A$-term which is related to the top quark Yukawa coupling, $A_t$. The 
reason is that we consider only $A_t$ as a relevant trilinear term for  
the electroweak radiative  breaking. Its expression is:
\begin{equation}
A_t= - m_{3/2}\ (\sqrt{3} \sin\theta - 3\cos\theta).
\label{a}
\end{equation}
For the gaugino masses $M_i$, taking the Green-Schwarz parameter 
$\delta_{GS}= -10$ \footnote{ $\delta_{GS}$ is shown to vary in the range 
$-5 \le  \delta_{GS} \le -10 $ for orbifold compactification \cite{Ibanez}.
Changing $\delta_{GS}$ in this range does not significantly affect the 
results of our analysis} we obtain:
\begin{equation}
\begin{array}{l} 
M_3= \sqrt{3}\ m_{3/2}\ (\sin\theta + 0.12 \cos\theta),\\ \nonumber
M_2= \sqrt{3}\ m_{3/2}\ (\sin\theta + 0.06 \cos\theta),\\ \nonumber
M_1= \sqrt{3}\ m_{3/2}\ (\sin\theta - 0.02 \cos\theta).
\label{gaugino}
\end{array}
\end{equation}

Finally we have to deal with the scalar bilinear soft breaking 
term $B \mu H_1 H_2$ (where $H_1$ and $H_2$ denote the scalar doublets), 
which strictly depends on the origin of the $\mu$-term in the 
superpotential. The smallness of $\mu$ in comparison with some typical 
superlarge scale (in our case the string scale) finds a natural 
explanation if $\mu$ arises solely from couplings in the K\"{a}hler 
potential~\cite{Giudice}. Since these couplings are indeed there 
in string theory 
it becomes appealing to view them as the only source of the 
$\mu$-term~\cite{Kaplunovsky}. In this case $B$ takes the form~\cite{Ibanez}:  
\begin{equation} 
B_Z = m_{3/2}\ (2+5 \cos\theta + 3\cos^2\theta).
\label{bz}
\end{equation}
A second option pointed out by ~\cite{Ibanez} is that $\mu$ arises 
solely from the $S$ and $T$ sector. Then:
\begin{equation}
B_{\mu} = m_{3/2}\ (-1-\sqrt{3} \sin\theta + 2\cos\theta).
\label{bmu}
\end{equation}
(We recall that, in the formula of $A_t$, $B_{\mu}$ and $B_Z$, we have 
used the above values of the modular weights).

Obviously it might well be the case that the mechanism 
originating $\mu$ is kind of admixture of the two above possibilities and 
then $B$ would be some combination of $B_{\mu}$ and $B_Z$ and we should 
consider it as an additional free parameter in the determination of the 
SUSY mass 
spectrum. For definiteness, 
in this work we will concentrate on the case of $B$ being 
$B_{\mu}$. This option for $B$ allows for a larger region of SUSY 
parameter space available for electroweak radiative breaking, although
it is maybe less attractive in the string context. A more general 
analysis including the $B_Z$ option as well as the case of $B$ as an 
additional independent parameter will be presented 
elsewhere~\cite{us}. 

Given the boundary conditions in equations \refe{scalar}, \refe{a},
\refe{gaugino}, \refe{bmu} at the 
compactification scale $M_S=3.6 \times 10^{17}$ GeV, we have to 
determine the 
evolution of the couplings according to their renormalization group 
equation (RGE) to finally compute the mass spectrum of the SUSY particles 
at the weak scale. In using the RGE's we keep only the top Yukawa 
coupling $\lambda_t$~\cite{Lopez}, \ie\ 
we assume that $ \lambda_t \gg \lambda_b, 
\lambda_{\tau}$. In so doing we are automatically leaving aside of our 
discussion those large values of $\tan \beta$ for which $\lambda_t \simeq 
\lambda_b$. We postpone the discussion of the large $\tan \beta$ regime 
to the abovementioned longer analysis~\cite{us}.

First we impose the condition of electroweak symmetry breaking. 
The potential for the two neutral components $H_1^0$ and $H_2^0$ of the 
higgs doublets reads~\cite{Nilles}:
\begin{equation}
\begin{array}{rl}
V=& (m_{H_1}^2 + \mu^2)\ \vert H_1^0 \vert^2 
+ (m_{H_2}^2 + \mu^2)\ \vert H_2^0 \vert^2 
- B \mu\ (H_1^0\  H_2^0 +{\rm h.c.})\ +\\[1ex] \nonumber
&\displaystyle 
\frac{1}{8}\ (g_1^2+g_2^2)\ (\vert H_1^0 \vert^2 - \vert H_2^0 \vert^2 ). 
\end{array}
\end{equation}
($m_{H_1}^2$ and $m_{H_2}^2$ satisfy the boundary condition at \ms\ 
\refe{scalar}, with the modular weights for $H_1$ and 
$H_2$ as in eq.\ \refe{modular}; we recall that
the product $B\mu$ can be assumed to be non-negative by
appropriate choice of the Higgs field phases). 
As usual the electroweak 
symmetry breaking requires the following conditions among the 
renormalized quantities: 
\begin{equation}
\begin{array}{l}
m_{H_1}^2 + m_{H_2}^2 + 2\ \mu^2 > 2 B \mu, \\ \nonumber
(m_{H_1}^2 + \mu^2)\ (m_{H_2}^2 + \mu^2) < (B \mu)^2,
\end{array}
\end{equation}
and
\begin{equation}
\begin{array}{l}
\mu^2 = \displaystyle 
\frac{ m_{H_1}^2 -m_{H_2}^2 \tan^2\beta}{\tan^2\beta - 1} 
- \frac{M_Z^2}{2},\\[1ex] \nonumber
\sin 2\beta= 
\displaystyle \frac{2  B \mu }{m_{H_1}^2+m_{H_2}^2+ 2\mu^2 },
\end{array}
\label{minimization}
\end{equation}
where $\tan\beta= 
{{\langle H_2^0 \rangle}/{\langle H_1^0 \rangle}}$.
A further constraint on the parameter space is entailed by the demand of
colour and electric charge conservation. In particular, the latter 
conservation yields a powerful constraint \cite{Savoy}, that has
been taken into account in our analysis.

Since $\mu^2$ and $\tan\beta$ can be expressed 
in terms of $m_{H_1}^2,$ $m_{H_2}^2$ and $B,$ all 
quantities depend in last analysis on the boundary conditions
at \ms. To be sure, in the evolution of the parameters entering 
equation \refe{minimization} there is also a dependence on the top Yukawa 
coupling. However we take the mass of the top, $m_t=174$ GeV\footnote{A 
discussion of the dependence of our results on the experimental error in 
the determination of $m_t$ will be provided in \cite{us}.}, as an 
experimental input and, then, using $\lambda_t={m_t}/(v \sin\beta),$
with  $v=\sqrt{{\langle H_1^0 \rangle ^2 
+ \langle H_2^0 \rangle ^2}}$ 
$=174$ GeV,
we can express $\lambda_t$ in terms of $\sin\beta$. In conclusion all low 
energy quantities are just functions of \mg, \th\ and sign of $\mu$. 
Obviously if instead of considering the definite situation for $B$ in which 
$B=B_{\mu}$ or $ B=B_Z$ one does not make any  commitment on $B$, then also 
this parameter should be added to the above  list. As we said, in the 
present analysis we consider only the case of $B=B_{\mu}$. For actual 
computation we take into account also the  one-loop 
corrections to the scalar potential due to the top-stop exchange which 
are known to affect the Higgs masses in a relevant way.

We now come to the main bulk of our analysis. The allowed region in the 
parameter space has to satisfy the usual requirement that chargino and 
sfermion masses are $\geq M_Z / 2$
and we further demand that the lightest SUSY particle (LSP) be a neutralino
or sneutrino, but should not carry electric or colour charge. Actually, this 
constraint is automatically fulfilled once electric charge conservation 
is implemented. As we 
mentioned above, we want to focus our analysis on the implications for \LEP2
physics. We will briefly denote by ``\LEP2\ available region" those points 
of the parameter space for which at least one of the charginos, sleptons 
or squarks is lighter than $M_Z$.

As we have seen, the whole low energy  spectrum is determined in terms 
of \mg\  and \th. In fig.\ 1 we show the \LEP2
available region in the (\th, \mg)  plane. The excluded area at the bottom 
part of the figure corresponds to points 
where some SUSY particle is too light. 
The vast dotted area occupying the center of the figure represents a region 
of the parameter space which is unavailable to \LEP2\ physics according 
to our previous definition. In conclusion the \LEP2\ available region which
is not already experimentally excluded corresponds to the blank area. 
The vertical solid line denotes the value of the goldstino angle which 
corresponds to the pure dilaton case $(\theta = \pi/2)$. 

A remarkable  feature of the figure is that in the pure or almost pure 
dilaton case the gravitino mass is constrained to be below 80 GeV or so 
to warrant \LEP2\ discovery of some SUSY particle, while rather higher 
value of \mg\ are available when we move to situations of significant 
admixture of $\langle F^S \rangle $ and $ \langle F^T 
\rangle $ contributions to the SUSY breaking. Also we can see that the 
available range for \th, where the correct  
electroweak breaking takes place  and all the experimental bounds on the SUSY 
particles are satisfied, is very limited (approximately 
$\theta \in [0.98,2]$ \rad\footnote{There is an analogous 
region in the neighborhood of $\theta=3\pi/2,$ for which 
an analogous discussion applies.}). 
These values of \th\  strictly depends on the 
values of modular weights given in equation \refe{modular}.  

The next relevant question becomes: which SUSY particle(s) is (are) most 
likely to be seen at \LEP2\ in the minimal string unification scenario that
we are studying here?  The answer is provided in figs.\ 2 and 3 where we plot
the values of the mass of the right-handed selectron and lightest 
chargino, respectively, 
as a function of the goldstino angle corresponding to the 
\LEP2\ available region in fig.\ 1 The dots constitute kind of 
iso-gravitino mass curves. The value of \mg\ increases going from the lower
to the upper part of the figure. The message that the two figures convey is
the following: considering the \LEP2\ available region, for non-negligible
dilaton-moduli admixture ($ \theta < 1.2 $ \rad) 
the mass of $ \tilde{e}_R $ is {\it always} within the \LEP2\ discovery 
reach, while 
for 1.2 \rad $< \theta < 2$ \rad, \ie\ for a case closer to the pure 
dilaton situation, the lightest chargino is {\it always} lighter than 90 
GeV.
Notice also that finding a right selectron with mass lighter than 70 GeV 
would be a signal for a departure from the pure dilaton scenario.
 
In fig.\ 4 we plot the mass of the lightest chargino vs. the mass of  
$\tilde{e}_R$  for the points of the \LEP2\ available region varying \mg\ 
and \th. 
The dilaton case corresponds to the highest allowed
values of $m_{\tilde e_R}$ for chargino masses in the range
$45$-$90$ GeV.
While there are several 
points corresponding to a lightest chargino mass $ > 90$ 
GeV, few points with $m_{\tilde{e}_R}> 90$ GeV are present. 
If a chargino is  seen at \LEP2\ it 
is very likely that also $\tilde{e}_R$ is 
visible there.

In fig.\ 5 and 6 we show the correlation between the mass of the lightest 
higgs and the mass of the lightest chargino  and $\tilde{e}_R$, 
respectively. From
fig.\ 5 we gather that the ``visibility" of the lightest chargino at \LEP2
implies that also the lightest higgs is visible. This does not hold true 
in the case of visibility of  $\tilde{e}_R$, since fig.\ 6 shows that for 
$m_{\tilde{e}_R} < 90 $ GeV there exist several points corresponding 
to a mass of 
the lightest higgs above 90 GeV. Indeed, it might be interesting to ask 
about the ``visibility" of the lightest higgs at \LEP2\ if some SUSY particle
(essentially, the $\tilde{e}_R$ or the lightest chargino, in our 
analysis) is discovered there. Fig.\ 7 shows that in correspondence to the 
``\LEP2\ available region" the mass of the lightest higgs is always below 
90 GeV  for the case of almost pure dilaton, while it can grow above 90 
GeV when 
$\theta < 1.2 $ \rad, \ie\ with conspicuous 
admixture of $\langle F^S \rangle $ and $\langle F^T \rangle $.
 
Finally, we notice that the sneutrino tends to be heavier than the 
$\tilde{e}_R$  or the lightest chargino. There exist vast areas of the 
parameter space 
region available at \LEP2\ where $\tilde{e}_R$ and/or the lightest chargino 
have mass $< 90 $ GeV, while the sneutrino mass is above 90 GeV.

In conclusion in this letter we have shown that schemes of minimal string 
unification provide interesting and rather detailed implications on 
physics to be tested in coming machines, in particular \LEP2. Clearly 
many results we reached in our analysis are tightly related to the choice 
of modular weights or the origin of the $\mu$-parameter or some other 
assumption we make. We plan to provide a more exhaustive analysis of the 
general phenomenological features of effective supergravities in a 
forthcoming work~\cite{us}. 
\vskip0.7truecm
\noindent{\Large\bf Acknowledgements}
\vskip0.5truecm
The authors would like to thank K. Benakli, S. Bertolini, L. Iba\~{n}ez, 
I. Lykken and K. Narain for useful discussions.
\vfill
\eject
\noindent{\Large\bf Figure Captions}
\vskip0.5truecm
\begin{description}
\item[{\bf Fig.\ 1}] The plane (\th,\ \mg).
The crossed region is excluded by direct mass searches;
the one with dots cannot be accessed by \LEP2.
The rest is the region which will be probed by \LEP2\ direct 
searches.
The solid vertical line \th$=\pi/2$ corresponds to the pure dilaton 
case.
\item[{\bf Fig.\ 2}] The right selectron mass in the 
\LEP2\ available region (see the text for definition) 
as a function of the goldstino angle.
The horizontal lines correspond to the visibility at \LEP2.
Notice, in comparison with next figure, the smaller vertical range.
\item[{\bf Fig.\ 3}] The lightest 
chargino mass in the \LEP2\ available region, 
as a function of the goldstino angle.
Horizontal lines as before.
\item[{\bf Fig.\ 4}] The right selectron mass versus the 
lightest chargino mass.
The vertical (horizontal) lines enclose the region in which
the chargino (right selectron) is visible at \LEP2\ energies.
\item[{\bf Fig.\ 5}] The lightest Higgs particle mass versus the 
lightest chargino mass. Vertical lines as before.
\item[{\bf Fig.\ 6}] The lightest Higgs particle mass versus the 
right selectron mass. Vertical lines correspond to visibility 
of right selectron at \LEP2.
\item[{\bf Fig.\ 7}] The lightest Higgs particle mass as a function of
the goldstino angle. The vertical line corresponds to pure dilaton case.
\end{description}
\vfill
\eject

\newpage

\pagestyle{empty}

\begin{figure}[t]
\psfig{figure=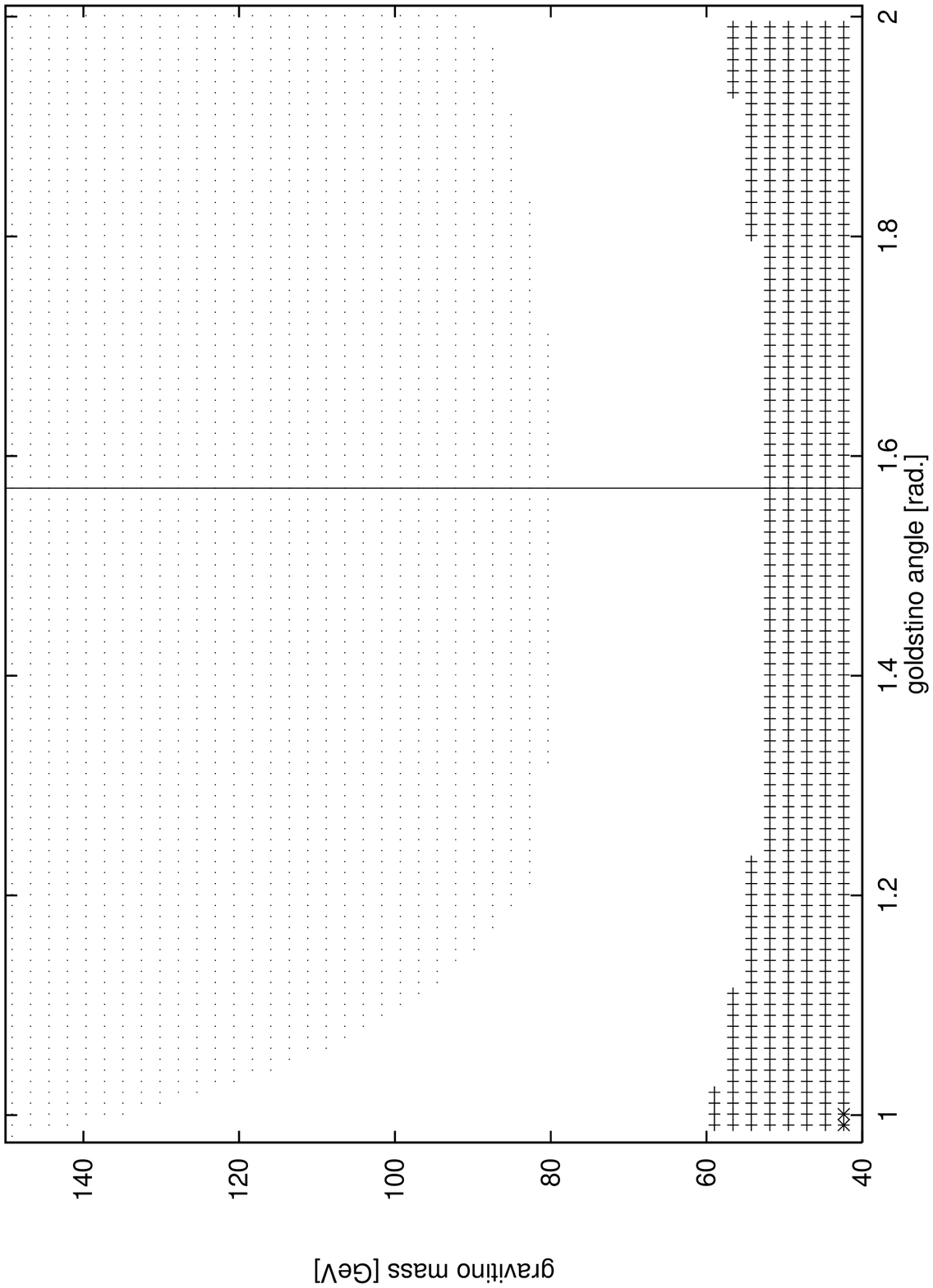,width=15.0cm,height=21.4286cm}
\end{figure}
\vfill\eject

\begin{figure}[t]
\psfig{figure=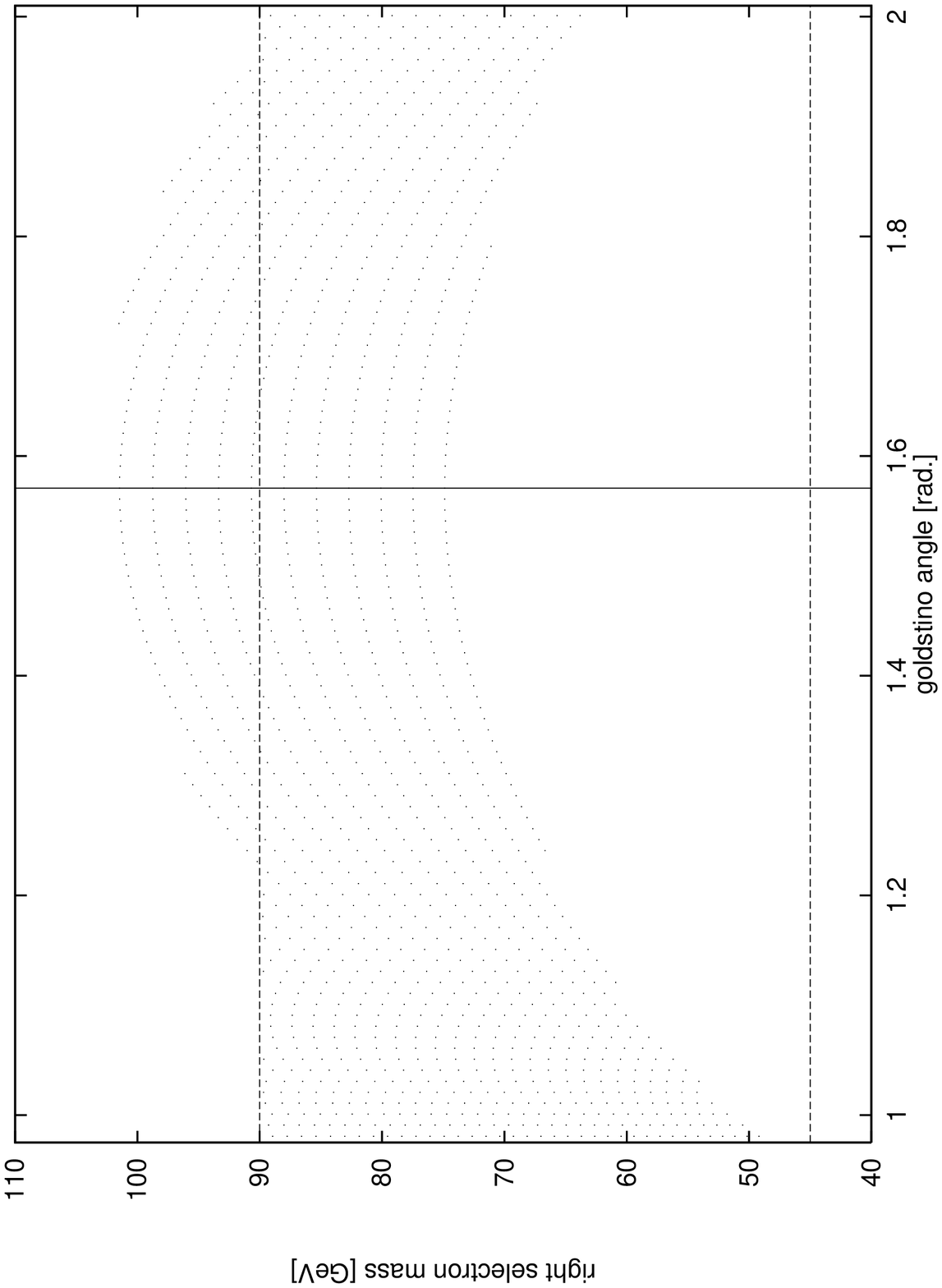,width=15.0cm,height=21.4286cm}
\end{figure}
\vfill\eject

\begin{figure}[t]
\psfig{figure=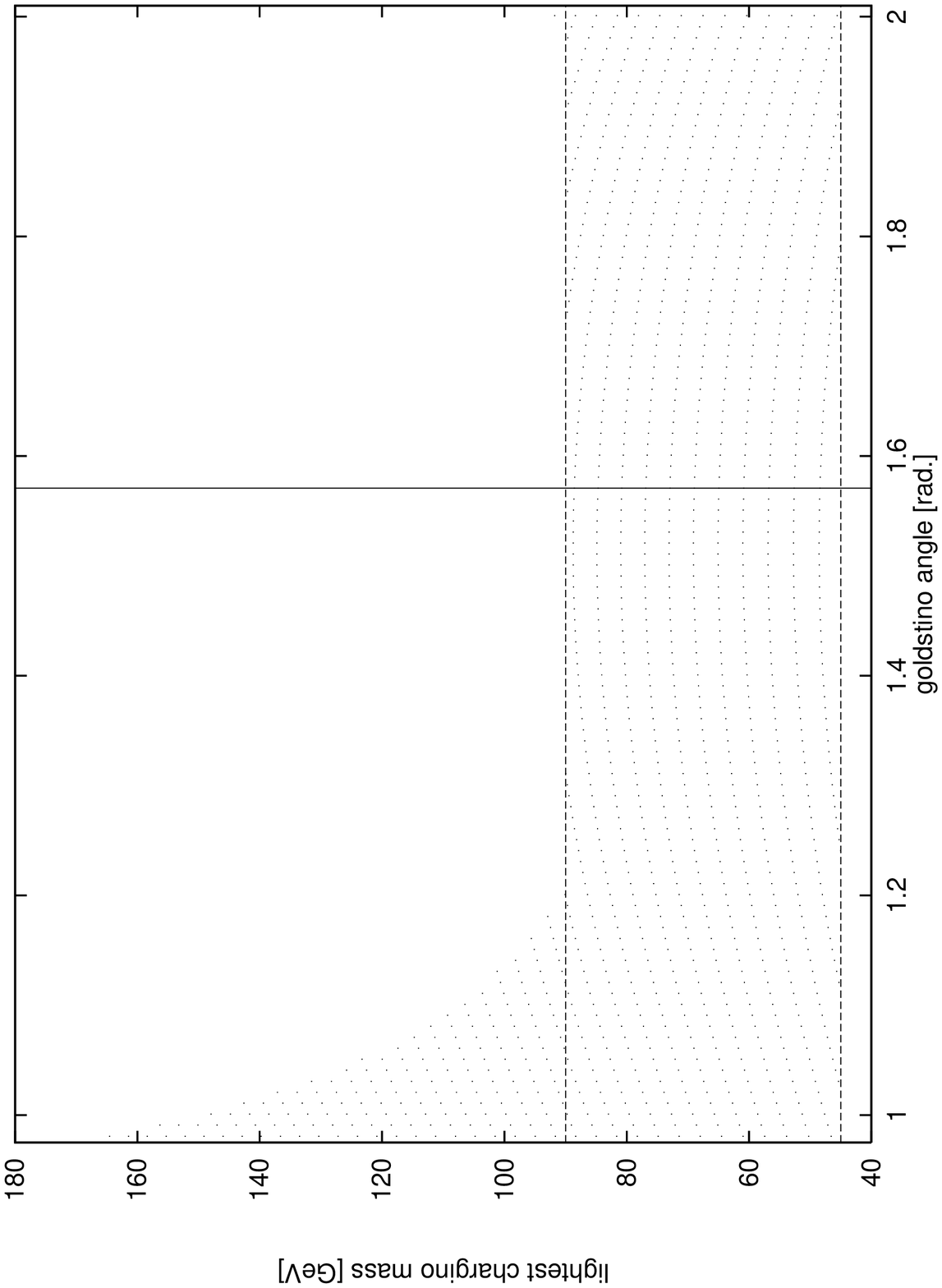,width=15.0cm,height=21.4286cm}
\end{figure}
\vfill\eject

\begin{figure}[t]
\psfig{figure=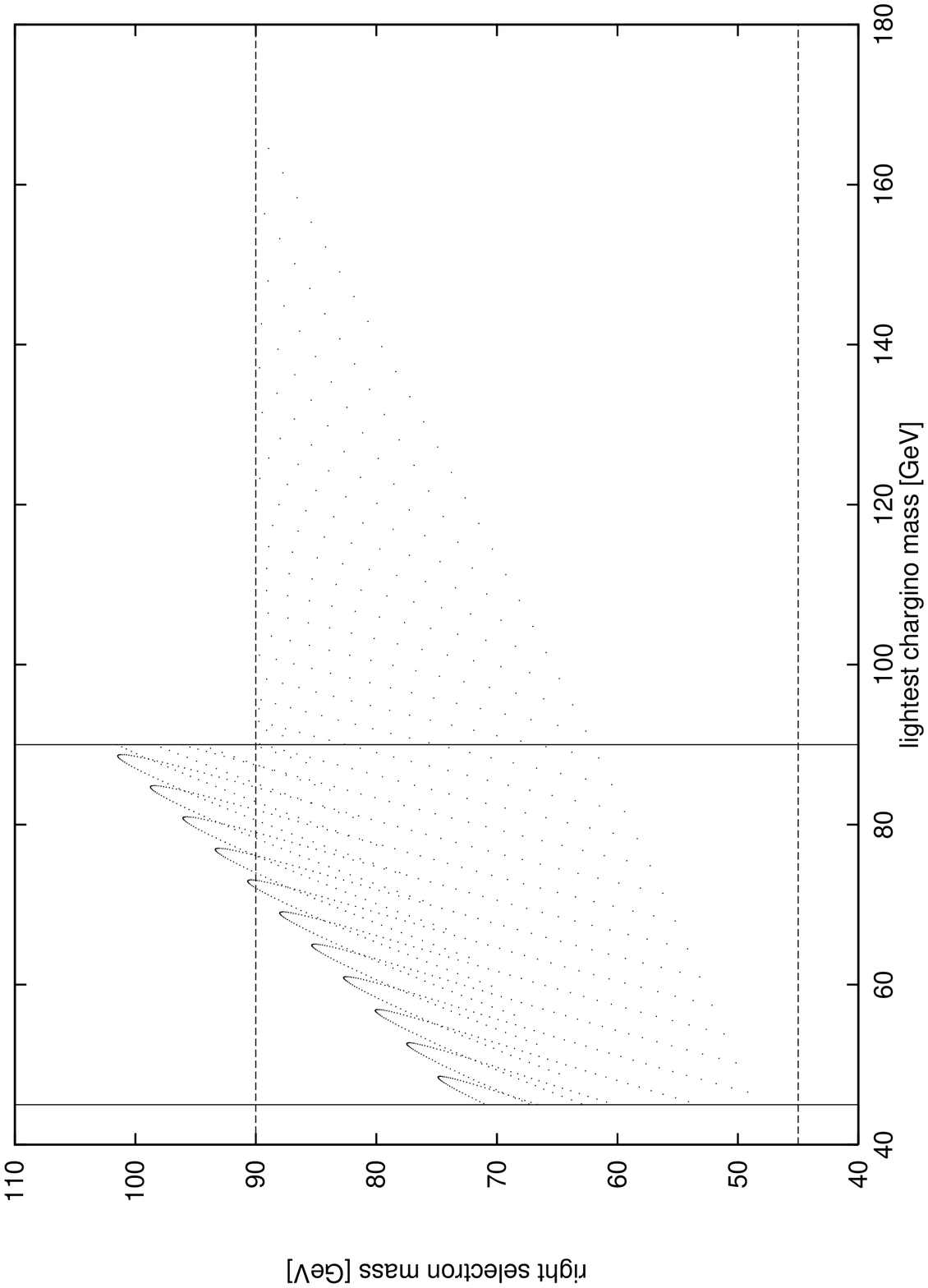,width=15.0cm,height=21.4286cm}
\end{figure}
\vfill\eject

\begin{figure}[t]
\psfig{figure=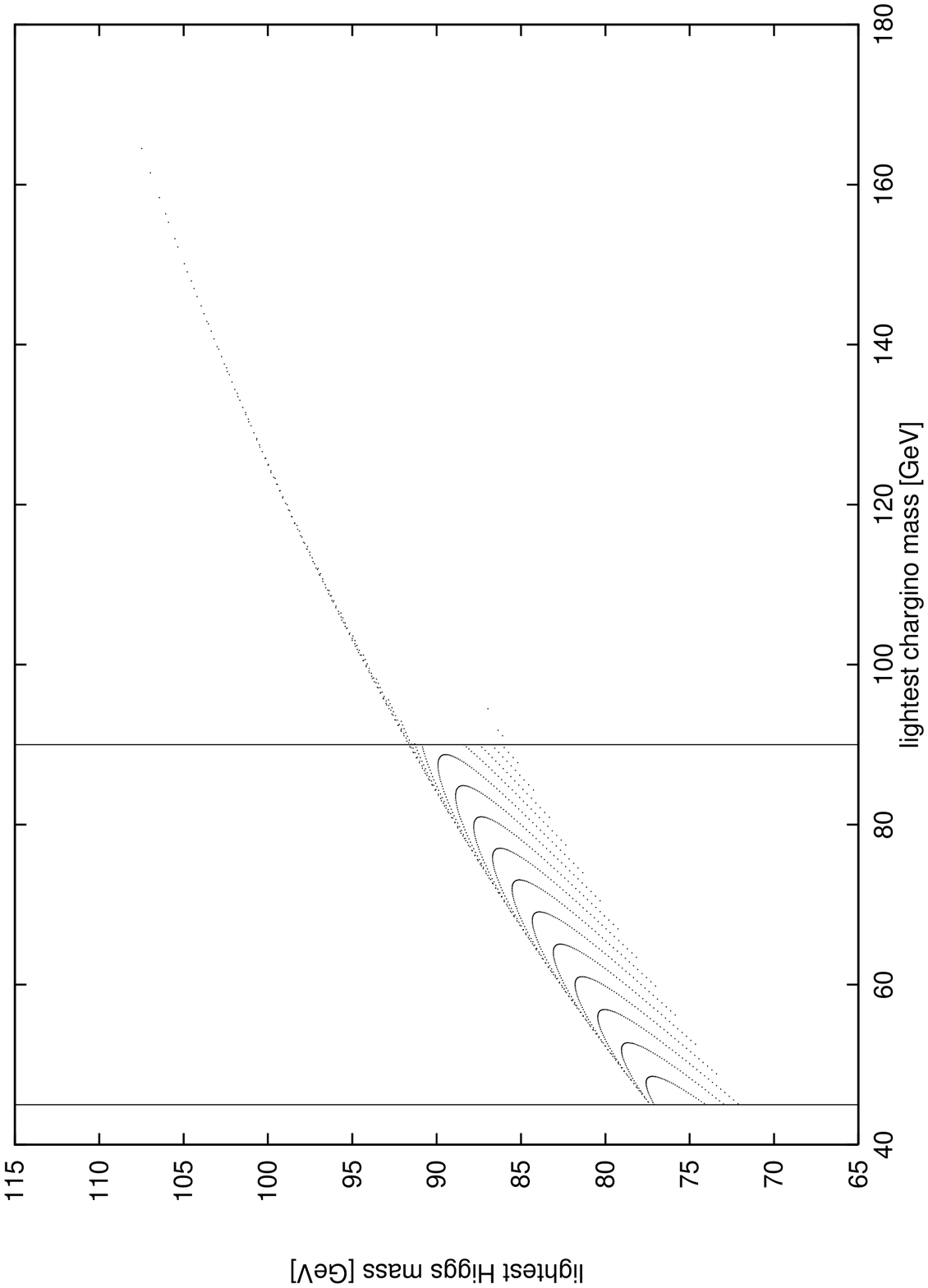,width=15.0cm,height=21.4286cm}
\end{figure}
\vfill\eject

\begin{figure}[t]
\psfig{figure=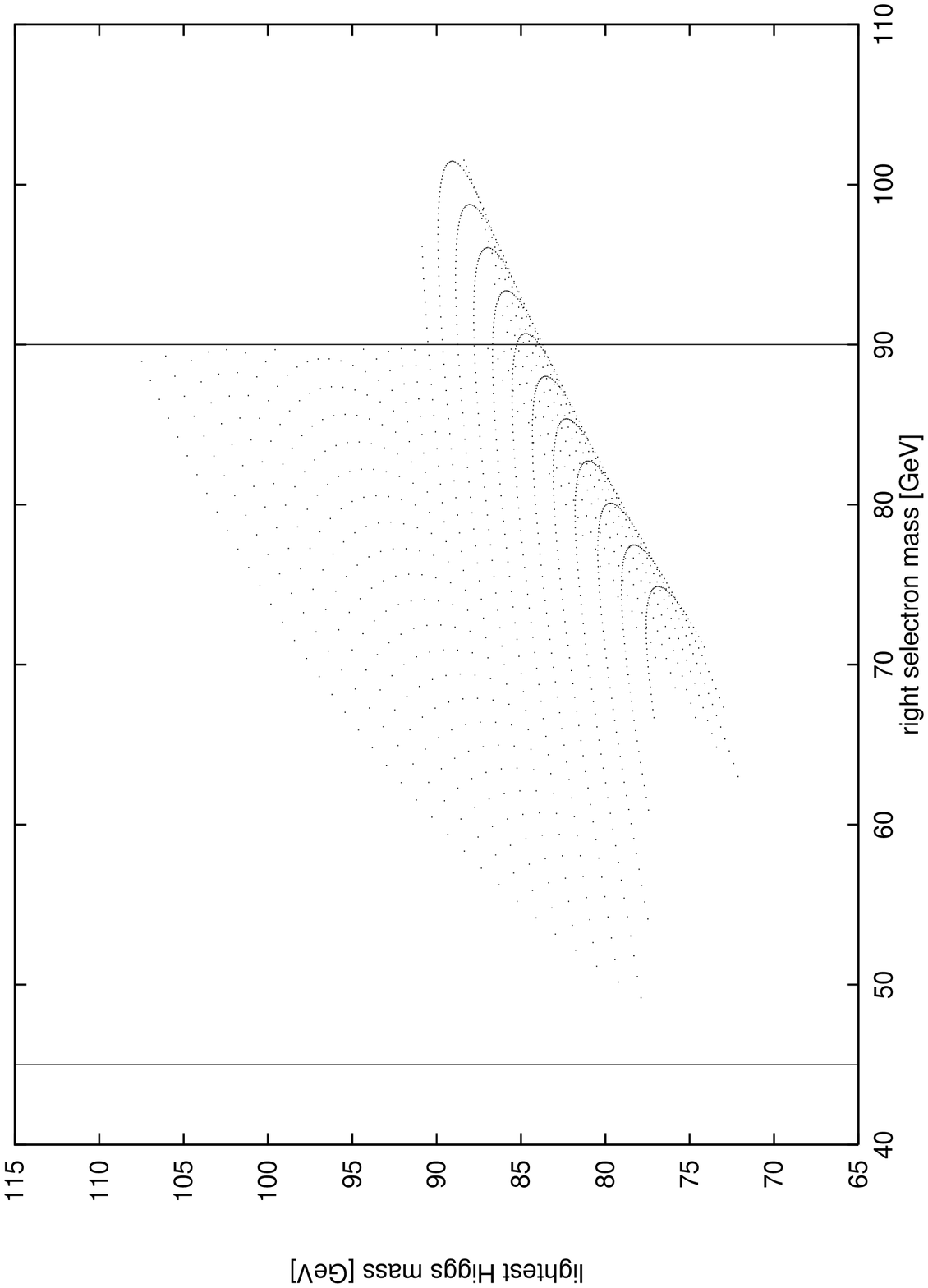,width=15.0cm,height=21.4286cm}
\end{figure}
\vfill\eject

\begin{figure}[t]
\psfig{figure=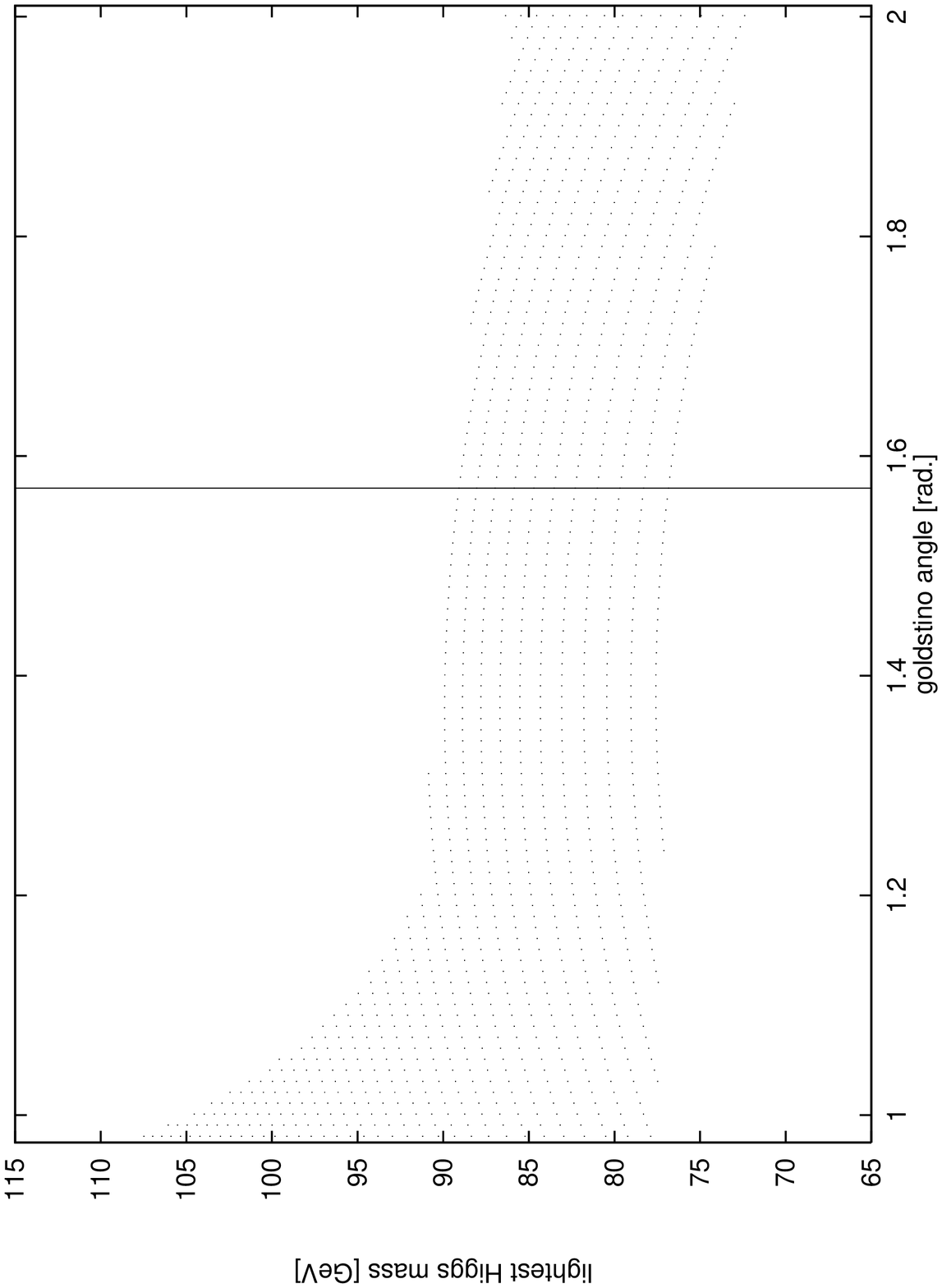,width=15.0cm,height=21.4286cm}
\end{figure}


\end{document}